# Electrically Switchable Broadband Photonic Bound States in the Continuum

*Andreas Henkel* [a,b], *Maik Meudt* [a,b], *Maximilian Buchmüller* [a,b], *and Patrick Görrn* [a,b*]

A. Henkel, M. Meudt, M. Buchmüller, and Prof. P. Görrn

[a] Chair of Large Area Optoelectronics, University of Wuppertal, Rainer-Gruenter-Str. 21, 42119 Wuppertal, Germany

[b] Wuppertal Center for Smart Materials & Systems, University of Wuppertal, Rainer-Gruenter-Str. 21, 42119 Wuppertal, Germany

*E-mail: goerrn@uni-wuppertal.de

**Abstract**

The question of how to continuously manipulate a photonic system between a radiating state and a bound state is an important challenge in photonics, as its solution promises broad technological relevance for optical sensors, modulators, switches and displays. Existing approaches utilise the inherent wave-nature of electromagnetic fields to their advantage, and are commonly identified as bound states in the continuum (BICs), as they resemble singular bound states embedded in a band of radiating states[1–24]. Although quasi-BICs have been demonstrated for numerous symmetric periodic photonic crystals, their existence so far has been limited to narrow spectral ranges, and their application in switches or modulators needs large changes of the refractive index[7,25].

Here, we show that the incidence of two guided symmetric substrate waves of opposite phase onto a symmetric periodic film waveguide enables the excitation of self-stabilising BICs which are electrically switchable in a broad optical range, with small changes of the refractive index. An experimental verification of the concept shows a switching contrast $C = 700$ at a wavelength of 532 nm and $C = 1000$ at a wavelength of 632.8 nm.

## Introduction

In the common understanding of physics, solutions of wave problems are separated into bound modes and radiative modes in non-overlapping frequency ranges. A peculiar deviation from this picture, initially described as the solution of a rather abstract artificial quantum mechanical potential by Neumann and Wigner[26], tells us that bound modes can be embedded within a sea of radiative waves. Such bound states in the continuum (BICs) are singularities that do exist for a large variety of wave-physical systems[27].

In electrodynamics, BICs are non-propagating (At-$\Gamma$-BIC) or propagating (Off-$\Gamma$-BIC) modes in symmetric photonic crystals, and they show game-changing properties for topologically-protected polarisation charges[13,22], vortex beams[9,11,12], sensors[28–30], modulators[5,7,8,10] and lasers[4,17]. The most prominent identifier for a BIC is an infinite radiative quality factor which can be continuously coupled to the radiation continuum by detuning the dielectric environment[2,3,6,10,14–16,25,31,32]. In state-of-the-art photonic systems, imperfections and finite sizes limit the radiative quality factor, and a true BIC always turns into a quasi-BIC[1]. Nonetheless, the achievable quality factors are extremely high, and even quasi-BICs should theoretically provide an ideal platform for photonic modulators or switches. However, their practical application for high-contrast local switching of radiation often remains difficult, as even detuned quasi-BICs keep too large radiative quality factors to sufficiently affect the coupling rate. A second drawback of state-of-the-art BICs is the fact that they are singularities, and hence only occur at singular positions with respect to energy and momentum. Hence, BIC switching could only be demonstrated within small spectral ranges and by introducing large changes of the refractive index $\Delta n \approx 2 \cdot 10^{-1}$ via photodoping or chemical processes[7,25].

For the approach presented in this work, we first look at the phenomenon of BICs from a different point of view. For a conventional resonant mode, a photonic crystal with periodicity $\Lambda$ leads to coupling of radiation from the photonic system's eigenmodes to room modes via the lattice momentum $|\vec{G}| = \frac{2\pi}{\Lambda}$. A BIC is then characterised by a vanishing interaction of an eigenmode with the lattice momentum. This is also true when the photonic crystal is realised as a grating layer.

A different way to minimise the impact of a grating in the centre of a waveguide is to exploit a node of an odd mode, which can be understood as a superposition of two waves with a relative phase shift of $\pi$. A film or grating placed at that position of minimum intensity shows minimum overlap with the mode. This approach was first used to minimise electrode absorption in organic lasers[33,34]. It was later suggested for grating couplers in light concentrators and switchable waveguides [35,36]. The major drawback, when compared to a BIC, is the inability of the node approach to completely switch off the grating impact inside a single resonator waveguide. On the other hand, it is compatible with broadband operation in symmetric setups and with tuning, as it keeps field symmetry around the node – even in a detuned asymmetric waveguide.

In this work, we combine a guided mode in an electro-optic substrate resonator with a second symmetric film resonator containing a grating in its centre. We observe that, for both even and odd film modes, the resulting combined modes, termed hybrid modes, can show both even and odd symmetry. BICs are found in the first case, that we also term hybrid node mode. In contrast to conventional waveguide odd modes, the interaction of hybrid node modes with the grating completely vanishes in a self-stabilising way. Under relative small local changes of the refractive index provided by the electro-optic effect of $LiTaO_3$[37], the hybrid modes can be switched into even modes with maximised interaction with the grating and thus strong coupling to room modes. This phenomenon occurs over a broad optical range and thus, for the first time, shows broadband switchability of BICs. The reported BICs are first excited via quasi-BICs, but during propagation they converge to true BICs without radiation losses and – in a loss-free system – without any propagation losses.

## Results & Discussion

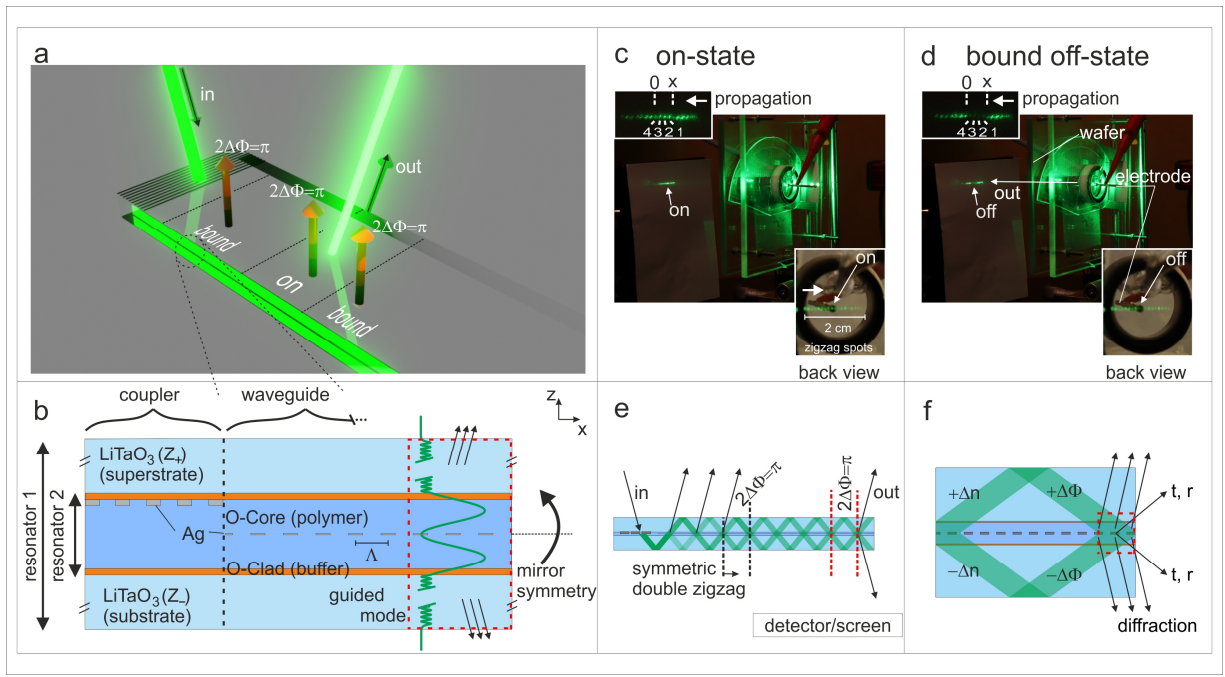

Figure 1. Operation principle of the switchable BIC-waveguide. a) Light is excited into the waveguide via a grating coupler. An internal phase shift of $2\Delta\Phi = \pi$ switches the propagating light between a BIC off state and a radiative on state. b) Geometry of the waveguide. It consists of a resonator 1, formed by the outer boundaries of two 0.5 mm thick LiTaO$_3$ wafers and a resonator 2 between the wafers, which is formed by a film waveguide consisting of two symmetric OrmoCore/OrmoClad stacks with a thin silver grating at its centre. For the excitation of light into the waveguide, a coupler region is used, for which the grating is positioned at one surface of the film resonator. c), d) Photographs of a locally-switched sample. The zigzag forms a line of radiating dots. The propagating light in the sample hits a phase-breaking defect at the position marked by "0". When an internal phase shift of $2\Delta\Phi = \pi$ is introduced at position "x", the diffracted radiation is cancelled until it reaches the defect. e) Level 1 of the model describing the waveguide: The light beam propagating inside resonator 1 experiences multiple reflection and transmission events at the film resonator and eventually splits into two zigzag paths with equal amplitudes. f) Level 2 of the model. Locally, the two zigzag paths excite leaky film modes. Due to the symmetry of the film resonator, perfect cancellation of the radiation diffracted into room modes occurs at a relative phase shift of $2\Delta\Phi = \pi$ between the two incident paths provided by the electro-optic effect. The opposite happens for $2\Delta\Phi = 0$ and diffracted radiation is emitted from the waveguide surface.

The operation principle of the waveguide is emphasised in Figure 1. At first, waveguide modes are excited using a laser and a grating coupler. The -1$^{st}$ order diffraction inside the waveguide can be controlled by an applied electric field symbolised by the red arrows, which induces an internal phase shift of $2\Delta\Phi = \pi$ (Figure 1a). In this way, the local emission is switched from the emission of a directed laser beam (radiative on-state) to an off-state (BIC) and vice versa. Figure 1b shows the geometry of the waveguide. It consists of two symmetric resonators. The first resonator (substrate resonator) is formed by two LiTaO$_3$ wafers of equal thickness $t_{LiTaO_3}$ with opposite orientations of the c-axis in growth direction (Z$_+$ and Z$_-$). The second resonator (film resonator) is formed by a film waveguide consisting of a thin rectangular metal grating symmetrically embedded between two polymer films (OrmoCore) with thin low-index polymer claddings (OrmoClad). Detailed parameters of the geometry can be found in the Methods section.

The switching behaviour of the device is emphasised in Figure 1c and Figure 1d. Figure 1c shows the on-state of the sample at a global value of $2\Delta\Phi = 0$. The dots are visible because the zigzag path of the substrate resonator interacts with the grating. Figure 1d shows the case of a local phase shift of $2\Delta\Phi = \pi$ at the position marked by an "x", which switches the light inside the waveguide into the BIC state, for which the dots appear dark. A phase-breaking event due to a local defect happens at another position marked by a "0", thus locally breaking the BIC condition and causing a radiative state. In this way, Figure 1d emphasises that the four dark spots are caused by a BIC still transporting the power of the laser beam. Tuning of the phase by applying a local electric field strength with a metal tip is well-suited for demonstrating local switching. However, this electrode geometry introduced optical scattering and a non-homogeneous electric field. That is why, in the following experiments, water electrodes were used instead (see Methods). With those electrodes a homogeneous external electric field strength $E_{ext}$ in the direction perpendicular to the waveguide surface is applied, and the refractive index of the top and bottom wafer is changed by $+\Delta n = \frac{1}{2}r_{13}n^3 E_{ext}$ and $-\Delta n$, respectively, in the present case of TE polarised waves[37]. Between the two crossing points of the zigzag, termed as dots, the upper and lower pathway thus experience a local relative phase-shift of $2\Delta\Phi = \frac{4\pi}{\lambda} 2 \frac{|\Delta n(E_{ext})| t_{LiTaO_3}}{\cos(\theta_{int})} [1]$.[38]

The functionalities of both resonators are explained in a local model with two hierarchical levels. The first level in Figure 1e emphasises the substrate resonator. After excitation into the waveguide, light propagates with an internal angle $\theta_{int} = \arcsin(\sin(\theta_{ext}) + \lambda/\Lambda)$ and experiences total internal reflection at the outer interfaces of the substrate resonator. Between each total internal reflection, the light is partially transmitted, reflected at the film resonator, and it forms two zigzag paths. With an increasing distance of propagation, the amplitudes of both zigzag paths level out to the same value. On the second level (Figure 1f), this can be locally considered as a symmetric two-beam excitation of the film resonator with field strengths $E_1$ and $E_2$.

So far, this mimics conditions similar to those of a simple two-way interferometer acting on the diffraction channels into room modes. However, the levelling of diffraction coefficients in amplitude and phase to achieve high values of $C$ is non-trivial, and has gained growing attention in the field of perfect coherent phenomena[39,40]. This is where the symmetry of the film resonator comes into play for broadband operation, as will be shown later on. To investigate this geometry experimentally, we fabricated a sample according to Figure 1b with a lamination procedure combined with a transfer printing process of the silver grating[41–43].

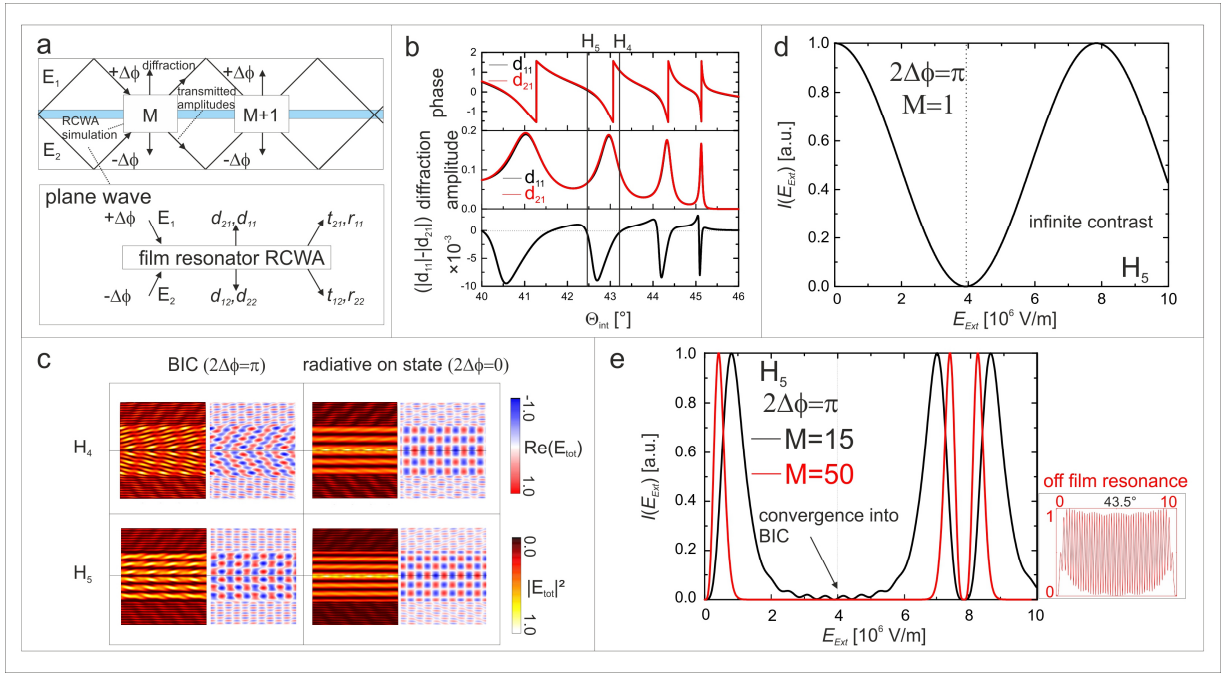

Figure 2. Simulations modelled by rigorous coupled-wave analysis (RCWA). a) Schematic of the simulation method. The film resonator is modelled by RCWA with two incident plane waves with a relative phase shift of $2\Delta\Phi$. b) Amplitudes and phases of the diffraction coefficients $d_{11}$ and $d_{21}$ as a function of the angle of incidence at 635 nm. Equal amplitudes $|d_{11}| - |d_{21}| = 0$ are reached nearby film modes. c) Cross-sectional plots of Re(E) and $|E|^2$ at the condition of equal coefficients for the $H_4$ and $H_5$ mode For $2\Delta\Phi = \pi$ between the plane waves (left column), an odd mode forms which avoids interaction with the grating. The opposite behaviour occurs for $2\Delta\Phi = 0$, for which the interaction with the grating is maximised. In the far field, these field distributions can be interpreted by perfect destructive (left column) and constructive (right column) interference which is mediated through the film modes. d) Outcoupled diffraction of the first dot in a line as a function of $E_{ext}$ for the $H_5$ mode. For $2\Delta\Phi(E_{ext}) = \pi$, the diffracted intensity reaches a value of zero and thus shows infinite contrast. e) Evolution of the outcoupled intensity $I(E_{ext})$ with a homogenous applied field strength $E_{ext}$ over increasing distance of propagation displayed as the number of dots $M$. Although many additive phase-shifts are collected along the propagating zigzag paths, $I(E_{ext})$ converges to a stable function when a film mode resonance is excited. For an infinite system, this convergence is a true BIC stabilised by film modes. Off film resonance, $I(E_{ext})$ does not converge (see inset for 43.5°).

To first explain the observed wave phenomenon at one single dot, we simulated its response to the two-beam excitation from the substrate via rigorous coupled wave analysis (RCWA) with a model of a film resonator, which was embedded in two infinite half-spaces of LiTaO$_3$ (see Figure 2a). In this model, the two paths of the zigzag are modelled as incident plane waves. Waves leaving the dot are then either passed to the diffracted intensity or to the further propagating zigzag beams. This is possible because the local modes of the film resonator are leaky to the substrate resonator and only show propagation lengths in the range of several tens of micrometres. Since the incoming beam has a width of approximately 1 mm, the film resonator can be approximated by a steady-state plane-wave solution. To investigate the response of film modes into diffracted radiation, the amplitudes and phases of the diffraction coefficients $d_{11}, d_{21}$ at a wavelength of 635 nm are shown as a function of the angle of incidence in Figure 2b. We observe that positions of equal amplitudes $|d_{11}| - |d_{21}| = 0$ (marked by vertical lines) are always close to the film modes. The near-field response of one dot to the two-beam excitation at the position of equal amplitudes is calculated via the superposition principle $\vec{E}_{tot} = \vec{F}(\vec{E}_1(\Delta\Phi)) + \vec{F}(\vec{E}_2(-\Delta\Phi))$, where $\vec{F}$ describes the electric field response of the film resonator to one incident plane wave. We term the resulting $\vec{E}_{tot}$ a hybrid mode $H_n$, whereby the index $n$ is named after the closest single-beam-incidence resonance $TE_n$ with $n$ nodes. Figure 2c shows $Re(\vec{E}_{tot})$ and $|\vec{E}_{tot}|^2$ for the two extreme cases of $2\Delta\Phi = \pi$ (left column) and $2\Delta\Phi = 0$ (right column) for the positions of equal amplitudes at $H_4$ and $H_5$ in Figure 2b. In the first case ($2\Delta\Phi = \pi$), we observe that $\vec{E}_{tot}$ is an odd field solution (hybrid node mode) for which the field intensity $|\vec{E}_{tot}|^2$ in the grating region is minimised. This is independent of whether the $H_4$ or $H_5$ mode is excited. The opposite happens for the second case ($2\Delta\Phi = 0$). Here, $\vec{E}_{tot}$ is an even field solution with maximised intensity $|\vec{E}_{tot}|^2$ inside the grating layers. Figure 2d shows how the diffracted intensity of a dot behaves as a function of $E_{ext}$ (see Figure 1f) for $H_5$. It is modelled by

$$I(E_{ext}) = |E_{out}(\Delta\Phi(E_{ext}))|^2 = |d_{11}(E_{ext})E_1 e^{i\Delta\Phi} + d_{21}(E_{ext})E_2 e^{-i\Delta\Phi} + d_{12}(E_{ext})E_1 e^{i\Delta\Phi} + d_{22}(E_{ext})E_2 e^{-i\Delta\Phi}|^2 \quad [2],$$

$C = \frac{I_{max}}{I_{min}}$ and the assumption of equal input amplitudes $E_1 = E_2$ (see Figure 1). For the later comparison with experiments, $\Delta\Phi$ is expressed through $E_{ext}$ via equation [1]. The diffraction intensity vanishes for $2\Delta\Phi = \pi$. If no external field is applied for the subsequent dots, the phase shift of $\pi$ is carried on and the vanishing diffraction remains stable. We further observe that the vanishing diffraction coincides with almost complete suppression of absorption. Figure 2e shows the evolution of the outcoupled diffraction $I(E_{ext})$ for the subsequent dots. $I(E_{ext})$ is calculated by passing the reflection and transmission coefficients $t$ and $r$ of the film resonator into the zigzag path to obtain the input amplitudes for the next dot. This procedure is then continued to receive the characteristic of the $M^{th}$ dot. Surprisingly, $I(E_{ext})$ converges to a stable function when a film resonance is excited. For $2\Delta\Phi = \pi$, the function converges to zero outcoupling for increasing values of $M$. At the same time, a broadening of the minimum around $2\Delta\Phi = \pi$ occurs. This convergence of the switching characteristics is independent of local disturbances such as mismatched input amplitudes or phases. On the contrary, if no film mode is excited, $I(E_{ext})$ does not converge to a stable function but instead shows an oscillating function with an increasing frequency on the field strength axis for increasing values of $M$ from the build-up of additive phases. We interpret this observation as the existence of a self-stabilising BIC, as radiation loss is completely suppressed when $M$ approaches infinity. This surpression of radiation is present for all diffraction orders and the convergence into a true BIC is also observed for loss-free gratings. The findings related to Figures 2d and 2e not only explain why, for the device at hand, a local relative phase-shift of $\pi$ turns the propagating wave from the radiative on-state into a BIC and vice versa, but they also explain why this type of BIC has not yet been found. An attempt to find it with conventional methods, by calculating the steady-state solution of a homogeneous infinite system excited by one plane wave, would only render a situation of global switching between the on and off state. The resulting steady-state solution then merely displays the average between both states, and the BIC remains hidden.

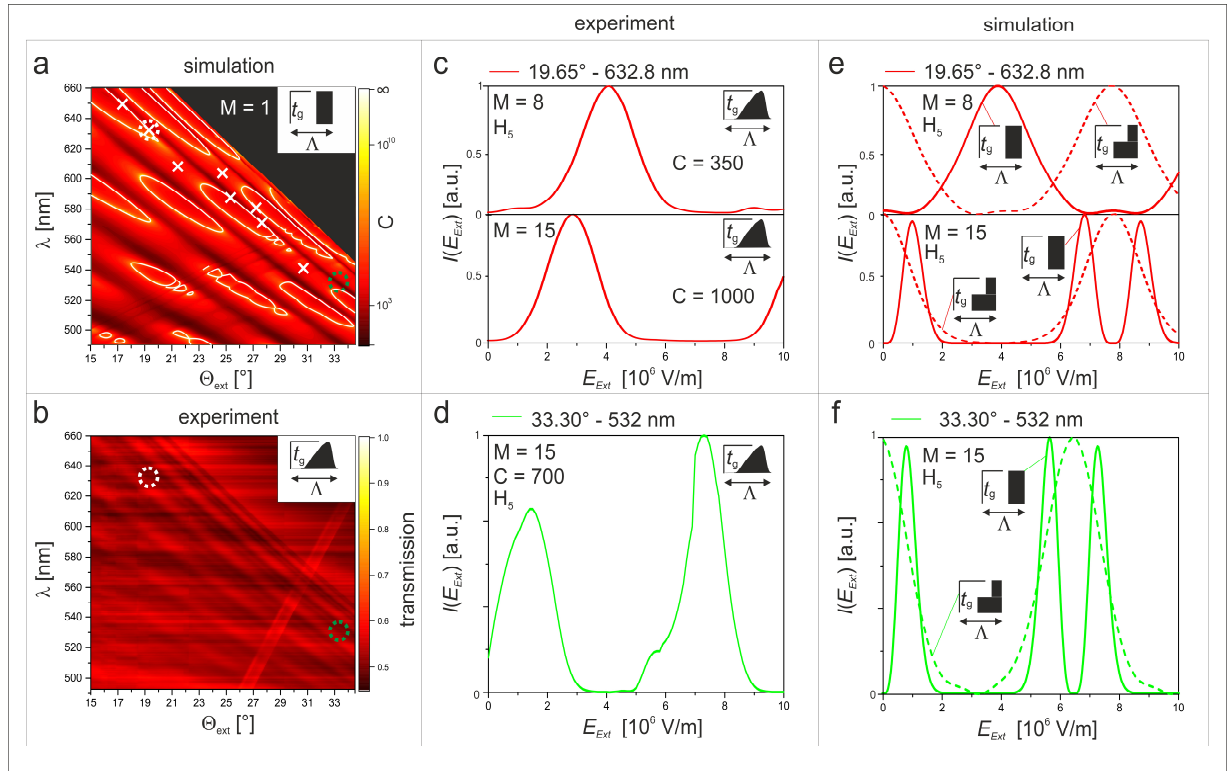

Figure 3. Broadband contrast and comparison to the experiment. a) Contrast map of a symmetric film resonator under two-beam excitation as a function of the wavelength and the outcoupling angle of diffracted radiation into air for a single dot ($M = 1$). Lines of infinite contrast occur along the film mode dispersion lines and imply the existence of BICs over a broad optical range. Convergence to BICs for infinite $M$ is confirmed for all positions marked by a white cross. b) Transmission measurement of the sample shown in Figure 1, for comparison. Transmission resonances occur at the predicted positions of film modes. c,d) Measured normalized diffracted power for M=8 and M=15 as a function of the external field strength for a wavelength of 632.8 nm, (white circle in 3a and 3b) and 532 nm (green circle in 3a and 3b). The insets show the grating profile within one period obtained from averaged atomic force micrographs. e,f) Simulated normalized diffraction for a fully symmetric waveguide (solid lines) and an asymmetric waveguide (slight grating and thickness asymmetry, dashed lines). Close agreement is observed between the calculated and experimentally-measured phase shifts as well as the broadening of the minima of the outcoupled diffraction with increasing values of $M$, as predicted in Figure 2e.

In Figure 3a, we show the simulated contrast for $M = 1$ as a function of $\lambda$ and the outcoupling angle $\theta_{ext}$. Please note that the outcoupling angle is normalised to air ($n = 1$) via Snell's law for a better comparison with experimental data. The contrast map consists of lines of high-contrast values, which follow the dispersion lines of film modes. Most strikingly, a broad coverage in the optical range by lines of infinite contrast is achieved for film modes of low $n$. This is due to the fact that the film resonator is symmetric, and bands of equal amplitude form along the dispersion relation of the film mode. For the positions marked by a white cross, we could confirm that convergence to a BIC is present for infinite values of $M$.

Figure 3b shows the results of an angular resolved transmission spectroscopy. Transmission resonances are observed at the expected positions of $TE_n$ film modes under single-beam excitation. Figures 3c and 3d show exemplary measurements of the outcoupled diffraction as a function of $E_{ext}$ at wavelengths of 532 nm and 632.8 nm. The measurement results of the 8th and 15th dot in a line are presented. The insets show the grating profile within one period obtained by averaged atomic force micrographs (AFM). The outcoupled intensity can be tuned within a maximal contrast ratio of $C = 1000$ for $\lambda = 632.8$ nm and $C = 700$ for $\lambda = 532$ nm for the $H_5$ mode. The periodicity of the signal matches perfectly the theoretically-calculated values of $\Delta\Phi(E_{ext})$ for M=8 and M=15. The corresponding results of the simulation are shown in Figure 3e and f. If a symmetric grating and layer thickness is assumed, deviations to the experiments are visible in the form of additional peaks for $M = 15$ around $8 \cdot 10^6$ V/m for $\lambda = 632.8$ nm and $6 \cdot 10^6$ V/m for $\lambda = 532$ nm, respectively (solid lines). This is presumably due to a slight asymmetry of both the grating layer and the resonator 1 as at least slight grating asymmetry and a film thickness fabrication tolerance of 0.4% is indeed present in the experiment (see Figure 3c and Figure 3d). This is confirmed by a toy model of a simple two layer asymmetric grating and a slight thickness modification of 0.4% per OrmoCore layer (see dashed lines). The assumption of such imperfection can explain the measured characteristics, and, importantly, confirms the robustness of the BICs. The experimental observations of increasing contrast and broadening of the off-state with growing $M$ agree with the model described in Figure 2, and thus confirm the existence of switchable BICs.

The long-desired goal of electrical high-contrast switching of laser emission from a slab waveguide has been achieved. Controlling the position of the directed laser emission can easily be translated to controlling its direction and shape as well, e.g., using lenses or curvy waveguides.

By exploiting switchable large bandwidth BICs, the geometric parameters of a light beam can be controlled by fast electro-optical phenomena. Beyond the obvious application in ultrafast laser displays, projectors and scanners, we anticipate that this new skill will eventually replace all mechanical systems that are needed today for light steering or tracking e.g., in concentrated photovoltaics and optical data storage.

In summary, we have shown a double resonator structure enabling photonic BICs of unsurpassed bandwidth that can be switched by an electric field. The structure consists of an inner leaky symmetric film waveguide with a central grating inside an outer second $LiTaO_3$ waveguide of slight asymmetry, which was modelled by a two-beam zigzag path with a tunable relative phase shift. The BICs form and self-stabilise for infinite propagation lengths in the vicinity of modes of the film waveguide, and are robust to non-ideal excitation conditions of the two-beam zigzag. Interestingly, even a single phase-shifting event, giving a relative phase shift of $\pi$ between the two beams, can lead to a quasi-BIC with both suppressed absorption and diffraction and thus there is no requirement of large system sizes to provide quasi-BICs. These findings render a different type of BIC in comparison with the conventional case of BICs from single beam excitation in homogeneous photonic crystals. We further consider these results as an important step to inspire broadband devices for perfect coherent phenomena as complete switching between a perfect zero and an enhanced on-state was shown. We anticipate that this new ultrafast geometric control of light beams will be a starting point for the next generation of optical systems needing no mechanical components.

## Methods

### Fabrication

The device stack (Figure 1b) was fabricated by laminating two pieces of a coated c-oriented LiTaO$_3$ wafer and, thus, enforcing mirror symmetry. To obtain this, the whole wafer $t_{\text{LiTaO3}} = 500$ µm was coated with a layer of OrmoClad $t_{OrmoClad} = 70$ nm and fully crosslinked by a UV curing step. On the upper half of the wafer, as suggested by Figure 1a and b, a line-shaped silver grating coupler with a thickness of $t_{\text{coupler}} = 100$ nm was added by transfer printing. After coating this stack with a layer of OrmoCore $t_{OrmoCore} = 1900$ nm and leaving it uncured, one half of the wafer was mechanically removed, UV-cured, and coated with a large area silver grating of a nominal thickness of $t_{\text{grating}} = 30$ nm by transfer printing. On the second half, the OrmoCore film was cured with about 15% of the full crosslink exposure dose, and so partially crosslinked. Subsequently, both wafer halves were laminated together in a lab press (p = 100 bar) and simultaneously UV-cured, forming a symmetric LiTaO$_3$/OrmoClad/OrmoCore/grating/OrmoCore/OrmoClad/LiTaO$_3$ stack. This procedure was developed in order to optimise the thickness homogeneity of all included films, as the reported phenomenon strongly relies on perfect symmetry.

Despite this effort, deviations of the measured characteristics (see Figures 3c and 3d, 3e and 3f) result from the asymmetric grating shape due to the transfer printing procedure or thickness inhomogeneities in the spin-coated films of up to ±0.4%. Remarkably, despite these variations, high-contrast switching is found on a substrate area as large as 10 cm², which is in agreement with the theoretical description and affirms the robustness of the presented approach.

**Simulation**

All calculations of the transmission and reflection coefficients as well as cross-sectional field amplitudes were performed by standard rigorous coupled wave analysis[44,45].

The experimentally-obtained diffraction efficiencies and propagation lengths, as well as reflection and transmission properties for various incident angles and wavelengths, were found to be approximately matched by modelling an ideal 30 nm thick rectangular grating, consisting of alternating blocks of silver and OrmoCore, with a unit-cell period of 555 nm and a filling factor of $FF = 0.35$ positioned at the centre of the OrmoCore film. For the two layer model to qualitatively describe the effect of an asymmetric grating shape, two 15 nm thick layers with filling factors of $FF = 0.7$ and $FF = 0.2$ were used, respectively. The remaining parameters are modelled as stated in the Fabrication section.

For the simulation of the theoretical voltage curves in Figures 3c and 3d, the equation for the outgoing diffracted intensity was calculated using

$$E_{out} = t_L \cdot (d_{11}e^{i\Delta\Phi} + d_{21}e^{-i\Delta\Phi} + d_{11} \cdot r_L \cdot r_F \cdot e^{i\Delta\Phi_0} + d_{21} \cdot r_L \cdot r_F \cdot e^{i\Delta\Phi_0} + d_{12} \cdot r_L \cdot t_F \cdot e^{-i\Delta\Phi_0} + d_{22} \cdot r_L \cdot t_F \cdot e^{-i\Delta\Phi_0})$$

with $\Delta\Phi_0 = \frac{4\pi}{\lambda} 2|\Delta n(E_{ext})|t_{LiTaO_3}/\cos(\arcsin((n_{LiTaO_3}\sin(\theta_{int}) - \lambda/\Lambda)/n_{LiTaO_3}))$, $r_L$ and $t_L$ as reflection and transmission coefficients from the substrate water interface, $t_F$ and $r_F$ as transmission and reflection coefficients under an angle of $\theta_{int,0} = \arcsin((n_{LiTaO_3}\sin(\theta_{int}) - \lambda/\Lambda)/n_{LiTaO_3})$.

For the calculation of the outcoupled intensity of an $M^{th}$ dot along the propagating zigzag, the transmission and reflection coefficients of one dot $t_{12}$, $t_{21}$, $r_{11}$ and $r_{22}$ were used to calculate the field amplitudes $E_{upper,M}$ and $E_{lower,M}$, which leave into the upper and lower part of the two beam zigzag. The equations read

$$E_{upper,M} = E_{upper,M-1}r_{11}e^{i\Delta\Phi,M-1} + E_{lower,M-1}t_{21}e^{-i\Delta\Phi,M-1} \quad [2]$$

$$E_{lower,M} = E_{upper,M-1}t_{12}e^{i\Delta\Phi,M-1} + E_{lower,M-1}r_{22}e^{-i\Delta\Phi,M-1} \quad [3],$$

whereby the outcoupled diffraction $I(E_{ext})$ is calculated analogously to equation [2] by replacing $E_1$ by $E_{upper,M-1}$ and $E_2$ by $E_{lower,M-1}$.

The optical constants of OrmoCore and OrmoClad were modelled using the Cauchy formula

$$n(\lambda) = A + \frac{B}{\lambda^2} + \frac{C}{\lambda^4}$$

with $A_{OCore} = 1.533$, $B_{OCore} = 0.00617\ \mu m^2$ as well as $A_{OClad} = 1.512$, $B_{OClad} = 0.00615\ \mu m^2$, and $C = 0$.

The wavelength-dependent optical constants of silver and water as well as electro-optic coefficients of LiTaO$_3$ were used for all simulations[37,46,47].

**Electrically-controlled output power measurement**

The electrically-controlled output power from the waveguide surface was always measured through a water-top electrode. Therefore, its impact on the guided modes and the room modes was minimised. The water was sealed against the sample using a rubber ring. Sample and rubber rings were mounted between two PMMA-plates.

For the bottom electrode, two different geometries were applied. A single metal tip covered with conductive silver paste was used to locally switch the zigzag beam phases. It was mounted in one of the PMMA-plates and insulated by silicone oil (see Figures 1c and 1d). In this way, we could switch off a line of dots, as may be seen in Figure 1d.

However, to characterise the sample in a more controlled way, a second set of experiments was conducted using a water-bottom electrode identical to the top electrode. In this way, a homogeneous global electric field strength over a length of 2 cm was applied, and the simulation procedure as shown in Figure 2a was used to theoretically model the outcoupled diffraction. We then attached a slit aperture to a power meter (Thorlabs, PM100USB, S150C) to measure the outcoupled diffraction of individual dots. To exclude any influence of stray light, a noise reduction tube was used, and the incoming laser beam was referenced by a second power meter (Thorlabs, PM100USB, S151C). The setup, consisting of electrodes and the sample, was mounted on an angular stage with an accuracy of 0.01°.

**Angular resolved transmission measurements**

To obtain angular resolved measurements, the sample was fixed on a two-axis rotational stage and a collimated white LED was used as an illumination signal. To compensate the lateral beam displacement of the 1 mm thick sample, the beam was reflected back on exactly the same pathway through the sample and guided to a grating spectrometer. The spectrum was measured with an angular resolution of 0.05° and a spectral resolution of $\lambda$ = 0.13 nm.

## Acknowledgement

This project has received funding from the European Research Council (ERC) under the European Union's Horizon 2020 research and innovation programme (Grant Agreement No. 637367). We gratefully acknowledge the European Regional Development Fund (ERDF) and the federal state of North Rhine-Westphalia (Grant No. EFRE-0400134), as well as the Federal Ministry of Education and Research (Photonics Research Germany funding programme, Contract No. 13N15390) for financial support.

## Competing Interests